\documentclass[12pt]{article}
\usepackage[dvips]{epsfig}

\def \ps
{p\!\!\!/}
\newcommand{\beq}{\begin{equation}}
\newcommand{\eeq}{\end{equation}}

\begin{document}

\title{Quadratic Effective Action for QED in $D=2,3$ Dimensions}
\author{D. Dalmazi\footnote{dalmazi@feg.unesp.br \, (On leave from UNESP-
Guaratinguet\'a - Brazil ).}\\ 
{\it Departament of Physics and Astronomy - SUNY} \\
   {\it  Stony Brook,  NY 11794 , USA }   \\
A. de Souza Dutra\footnote{dutra@feg.unesp.br} and 
Marcelo Hott\footnote{hott@feg.unesp.br} \\ 
{\it UNESP - Campus de Guaratinguet\'a - DFQ} \\
{\it Av. Dr. Ariberto Pereira da Cunha, 333} \\
{\it CEP 12500-000 - Guaratinguet\'a - SP - Brazil. } }
\maketitle
\begin{abstract}
We calculate the effective
action for Quantum Electrodynamics (QED) in $D=2,3
$ dimensions at 
the quadratic approximation in the gauge fields.
We analyse the
analytic structure of the corresponding nonlocal boson
propagators
nonperturbatively in $k/m$. In two dimensions for any nonzero
fermion
mass , we end up with one massless pole for the gauge boson . We
also
calculate in $D=2$ the effective potential between two static
charges
separated by a distance $\,L\, $ and find it to be a linearly
increasing
function of $\,L\,$ in agreement with the bosonized theory (massive
Sine-Gordon model). In three dimensions we find 
nonperturbatively
in 
$\, k/m \, $ one massive pole in the
effective bosonic action
leading to screening. Fitting the numerical results
we derive a simple
expression for the functional dependence of the boson
mass upon the
dimensionless parameter $\,e^{2}/m\,$
.
\\
{\it PACS-No.:} 11.15.Bt , 11.15.-q
\end{abstract}

\newpage

\section{Introduction}

The
mapping of fermionic theories into bosonic ones is a very
powerful
technique used to understand nonperturbative aspects of
quantum field theories. This so called bosonization is exact
in two
dimensional theories, and it has been widely employed in this
dimension [1-3]. In the last few years, many papers have
been devoted to the
study of what has been called bosonization in three
dimensions [4-8], and
even in four dimensions \cite{marino1}. This kind of
path-integral
bosonization consists of obtaining the effective action by
integrating out
the fermion degrees of freedom and then studying the
physical properties of
the resulting effective theory. In most of the
works along this line one
makes use of the derivative expansion
\cite{ait1, shif} and derives weak
bosonization rules for the system
\cite{schaposnik, banerjee, elcio} as well
as the particle content and
their masses \cite{schaposnik}. On the
other hand in \cite{barci} an
exact, in $k/m$, action at one-loop level is
used to show that the
bosonization of \cite{marino} and \cite{schaposnik2},
is recovered in the
large and small momentum limits respectively. In addition ,
the authors
of \cite{elcio} have studied the assymptotic
properties of the bosonic
effective action associated with QED in three
dimensions, showing the
screening property of the effective potential
between two static
charges.

In this work we apply the effective action approach of
bosonization
on QED in two and three dimensions. 
The usefulness of the
two dimensional case lies on the fact
that $QED_{2}$ can be bosonized via
the massive Sine-Gordon
 model which exhibits confinement
\cite{livroElcio, Jackiw, Coleman}.
Thus both approaches of bosonization can be
compared. Here we verify
that, indeed the confining behavior also
appears in the bosonization \textit{%
via} effective action. It is
remarkable
that in two dimensions at the 
quadratic approximation in the gauge fields but without any expansion in
$k/m$, the
massive pole of the Schwinger model disappears, being replaced
by a massless
pole, which is in agreement with what has been observed in
\cite{gross} , but
it differs from the result obtained through
perturbative ($\,m/e\,$)
calculation of \cite{adam}. In three dimensions
it is shown that there is a
massive excitation which depends on the
dimensionless parameter $\frac{16\pi
\,m}{e^{2}}$ . We have found a
simple approximated expression for this
function. This in fact
generalizes the calculations of \cite{schaposnik},
which were obtained at
leading order of the derivative expansion, and that
of \cite{cesar99}
carried out at a higher order in $k/m$, which in its turn
is related to
consistent higher derivative actions \cite{cesar97, jackiw99}.

In both
cases we consider the one-loop effective action up to second-order
in the
coupling constant $e$. We may write it
as

\begin{equation}
S^{(2)}_{eff}=- \frac{1}{2} \int \;
\frac{d^{D}k}{(2 \pi)^D} \; {\tilde{A}}%
_{\mu}(-k) \left[g^{\mu \nu} k^2
- k^{\mu} k^{\nu} - \Pi^{\mu \nu}(k)
\right]
{\tilde{A}}_{\nu}(k),
\end{equation}

\noindent where
${\tilde{A}}_{\mu}(k)$ is the Fourier transformation of $%
A_{\mu}(x)$
and

\begin{equation}
\Pi^{\mu \nu}(k) = i e^2 \int \; \frac{d^{D}p}{(2
\pi)^D} \; tr \left[ \frac{%
1}{\ps - m + i\epsilon} \gamma^{\mu}
\frac{1}{(p\!\!\!/ + k\!\!\!/) - m +
i\epsilon} \gamma^{\nu}
\right]
\end{equation}

\noindent is the polarization tensor obtained
after integrating out the
fermion fields. The space-time dimension is
represented by $D$ ($\, D=2,3\,$). It  is
useful to expand the polarization tensor
in powers of $\frac{k}{m}$ which
corresponds to the derivative expansion of
the effective action
\cite{das1}.
Truncating the expansion at some power of $\, k/m\,$ not only yields a
local effective
action but also allows one to analyse the r\^ole of some
specific term as
it is the case of the leading odd-parity term in
$QED_{3}$ \cite{das2}. It
is also important to study the phenomenological
aspects of a
low-energy effective action \cite{ait2}. The order at which the
series is
truncated depends on the range of energy one is
interested in.

Recent studies of the contributions of higher-order
derivative terms in a
low-energy effective gauge theory revealed the
possibility of the 
appearence of
non-physical excitations. Here we
overcome this difficulty by analyzing
directly the poles of the full
propagator at one-loop level up to second
order in the coupling
constant.

\section{Effective potential in $QED_2$}

In $QED_2$ the
polarization tensor will be given by

\begin{equation}
\Pi^{\mu \nu}(k)
= \Pi(k^2) \left( g^{\mu \nu} -\frac{k^{\mu} k^{\nu}}{k^2}
\right)
,
\end{equation}

\noindent where using , e. g. , dimensional regularisation we get

\begin{equation}
\Pi(k^2) =
\frac{e^2}{\pi} \left[ 1 + \frac{1}{2} \frac{4{m^2}/k^2}{(1
- {4{m^2}/k^2})^{1/2}}  \ln{\frac{(1- {4{m^2}/k^2})^{1/2} +1}{(1 -
{4{m^2}/k^2})^{1/2}-1}} \right] .
\end{equation}

We are interested in
the interacting potential between two static charges
$\, Q \, $ and $\, -
Q \, $ located at $x=L/2\,$ and $\, x=-L/2 \, $.
Solving\footnote{ We have not taken into account 
the general solution of the homogeneous differential 
equation ( without sources ) which amounts to
neglect the $\theta$-vacuum ($\, \theta =0 \, $)} the equation of motion
derived from the effective action
we obtain the potential produced
by the  positive
charge $Q$ : 
\begin{equation}
A_{\mu}(x) 
= \int \; \frac{d^{2}k}{(2 \pi)^2}\int
d^{2}x^{\prime} D_{\mu
\nu}(k) e^{i k (x - x^{\prime})}
J^{\nu}(x^{\prime}) ,
\end{equation}

\noindent
where

\begin{equation}
J^{\nu} (x^{\prime}) = Q
\delta({x_{1}}^{\prime} - {\frac{L }{2}})
\delta^{\nu
0}
\end{equation}
\label{current}

\noindent is the conserved current
and $D_{\mu \nu}(k)$ is the ``photon"
propagator whose longitudinal term
is given by

\begin{equation}
{D^\parallel}_{\mu \nu}(k) = \frac{1}{k^2
- \Pi(k^2)} g_{\mu \nu}.
\end{equation}

Due to the specific form of
the external current the only contribution to
the potential will be its
time-component . The corresponding integral
can be easily performed on
the complex plane for arbitrary masses . 
The resulting static potential
at $x_1=
- {\frac{L }{2}}$ grows linearly with the distance between the
charges,
namely

\begin{equation}
A_{0}(x=-\frac{L}{2}) = \frac{Q}{2
\left[1 + \frac{2}{3 \pi} (\frac{e}{2m}
)^{2} \right]}
L.
\end{equation}

Then we conclude that, at the quadratic 
approximation used here, $QED_2$ results
in a {\it confining} potential. 
Because of the quadratic 
approximation in the gauge fields 
we may say that the result obtained here,
{\it i.e.}, a linearly growing 
inter-fermion effective potential is due to a zero
mass pole for the gauge potential. One also reaches a similar result in the
usual bosonization  approach when the quadratic approximation
in the boson field of the massive Sine-Gordon model  
is used \cite{gross}. 
In other words we can say that, 
at the quadratic approximation, turning on a mass for the
fermion field 
will prevent the mass generation for
the gauge boson and the classical
result, that is a massless pole, prevails.
It is worth mentioning that 
, since we are only interested in 
the real contribution to the effective
action , we have dropped in (4) the
imaginary part of the polarization
tensor which appears 
beyond the pair creation threshold 
$k^2> 4m^2 $. More specifically ,
expression (4) is correct for $k^2<0$
which is the relevant region for 
the calculation of the effective potential
of static charges due to the factor
$\delta (k_0)$ which comes from the
time integral in (5). However,
for the analysis of the particle
content of (1)
we have  also calculated the polarization
tensor in the region $0<k^2<4m^2$  
where we found no poles except in
the limit $k^2\to 0^+$  
where we found an agreement with the limit
$k^2\to 0^-$ of (4).
Finally, it is important to notice that
although we can recover the Schwinger model
effective action from the $\,
m\to 0 \, $ limit of (4)
its static potential does not correspond to the
massless limit 
of (8) since the integral and the limit do not commute with each other. 
In the Schwinger
model (massless $QED_{2}$) a mass for the gauge boson is generated
by the gauge anomaly and as a consequence one obtains a
\textit{screening} potential between two static charges \cite{swieca, elcio2}.


\section{Massive pole in $QED_3$}

Here we are not concerned with the
explicit expression for the interaction
potential but with the behavior
of the mass generated dynamically as a
function of the coupling constant
and the fermion mass.

In this case the polarization tensor is given
by

\begin{equation}
\Pi^{\mu \nu}(k) = \Pi_{1}(k^2) i \epsilon^{\mu
\nu \rho} k_{\rho} +
\Pi_{2}(k^2) \left( k^{2} g^{\mu \nu} -k^{\mu}
k^{\nu} \right) ,
\end{equation}

\noindent
where 

\begin{equation}
\Pi_{1}(k^2) = - \frac{e^2}{8 \pi}
\frac{1}{\sqrt{k^{2}/ 4{m^2}}} \ln{\frac{%
1+ \sqrt{k^{2}/4{m^2}}}{1 -
\sqrt{k^{2}/ 4{m^2}}}}, \label{pi1}
\end{equation}

\noindent
and

\begin{equation}
\Pi_{2}(k^2) =  \frac{e^2}{16 \pi m}
\frac{1}{k^{2}/4{m^2}} \left[ 1 - \frac{%
1}{2} \left( \frac{1 +
{k^{2}/4{m^2}}}{\sqrt{k^{2}/4{m^2}}} \right) \ln{%
\frac{1+
\sqrt{k^{2}/4{m^2}}}{1 - \sqrt{k^{2}/ 4{m^2}}}}
\right]. \label{pi2}
\end{equation}

Equations (\ref{pi1}) and (\ref{pi2}) were obtained by using
dimensional 
regularization method. Equation  (\ref{pi1}) is regularization 
dependent and our result corresponds to an equal number
of Pauli-Villars regulators with
positive and negative masses. 
Like the two dimensional case ,
since  we are only interested in the real 
contribution to the effective
action,
we have dropped
the imaginary part of the polarization 
tensor which appears beyond the pair
creation threshold ( $k^2>4m^2 $ ).
More precisely we have only given the polarization
tensor (see (\ref{pi1}) and (\ref{pi2})) 
in the region $0<k^2<4m^2$ where we found a pole
in the photon propagator. For $k^2<0$
the correct result correspond to
replace $\frac{1}{\sqrt{u}}log((1+\sqrt{u})/(1-\sqrt{u}))$
by $\frac{-2}{\sqrt{-u}} arctan\frac{1}{\sqrt{-u}} \, $
in (\ref{pi1}) and (\ref{pi2}) where $u=\frac{k^2}{4m^2}$. 
The situation is similar to QED$_4$ , see , e.g., [26]. 
We have explicitly verified that no tachyonic poles appear
without any approximation on $\frac{k^2}{4m^2}$ .
Back to the region $0< k^2<4m^2 $ one can check that the ``photon"
propagator

\begin{equation}
{D^\parallel}_{\mu \nu}(k) = -
\frac{1}{\tilde{\Pi}_2 \left[ k^2 - (\frac{%
\Pi_1}{\tilde{\Pi}_2})^{2}
\right]} g_{\mu \nu},
\end{equation}

\noindent
where

\[
\tilde{\Pi}_{2}(k) = 1 - \Pi_{2}(k), 
\]

\noindent
develops a massive pole $k^{2}=M^{2}$. The gauge boson mass $M$
depends
on the coupling constant and fermion mass. We have carried out
a
numerical analysis for the behavior of this mass generated dynamically
and
found a very simple function ,
namely

\begin{equation}
M=\frac{2\,m}{c_{1}\,+\,c_{2}\,a\,+\,c_{3}\,a^{2}},
\end{equation}

\noindent
where

\begin{equation}
a=16\pi
\frac{m}{e^{2}}.
\end{equation}

\begin{figure}[tbp]
\begin{center}
\begin{minipage}{0.4\linewidth}
\epsfig{file=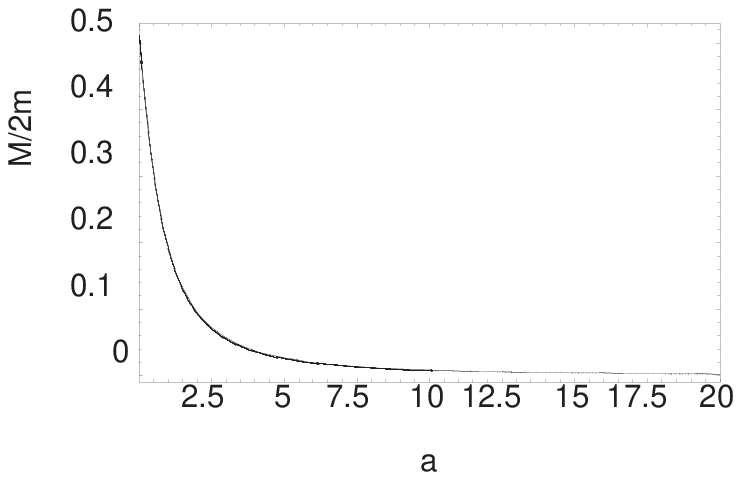}
\end{minipage}
\end{center}
\caption{Dependence
of the rate of the massive bosonic pole and 
fermion mass with respect to
the inverse coupling $\,a=16\pi \frac{m}{e^2} \,$.}
\label{fig:figure1}
\end{figure}

\begin{figure}[tbp]
\begin{center}
\begin{minipage}{0.4\linewidth}
\epsfig{file=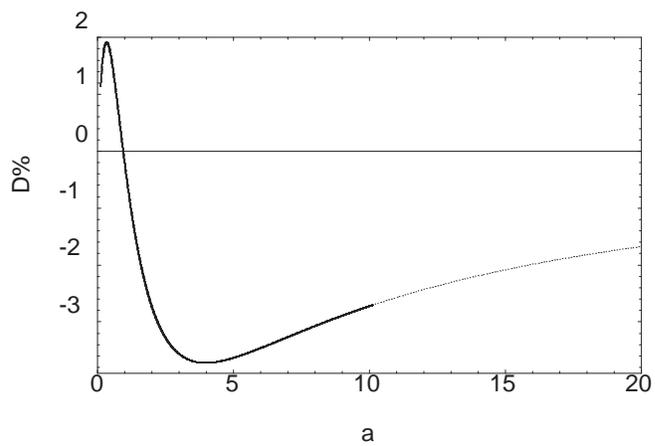}
\end{minipage}
\end{center}
\caption{Percentual
difference between the fitting and the numerical
data}
\label{fig:figure2}
\end{figure}

\noindent From the numerical results
for  the massive pole at small $a$ we get 
$%
c_{1}=1.953331381$ . On
the other hand , from the large  
$a$ region we have $c_{3}=1$.
Finally
the constant $c_2$ was adjusted by choosing the best fitting for a
curve
with about eleven thousand numerically calculated points, and it
was
found to be $c_{2}=2.253$. The maximum error of the fitted function
is less
than $3.8\%$, with the statistical controlling parameter
$\chi
^{2}=5.5\times 10^{-6}$. In the figure \ref{fig:figure1} we present
the
curve of the massive pole as a function of the dimensionless
parameter $a$.
Note that it is impossible to distinguish the exact from
the adjusted curve
in figure 1. For this reason we present in the figure
\ref{fig:figure2} the
dependence of the percentual deviation of the
fitted curve from the numerically
obtained masses.

In reference
\cite{elcio} the derivative expansion up to second-order in $%
\frac{k
}{m}$ is used to compute the effective potential which was also found  
to
be of the \textit{screening} type. We have taken into account the
whole
expression for the polarization tensor and found only one massive
pole 
in the whole range of the parameter $a$.  Since the pole can never
be found at
the origin we conclude that, nonperturbatively in $k/m\,$ the screening
effect prevails in agreement with the numerical analysis
carried out in
reference \cite{elcio3} for the static potential.
The
truncation based on the derivative expansion only amounts to a
displacement
of the massive pole from its nonperturbative ( in $k/m$ )
value.

\section{Acknowledgements}

We would like to thank professors
E. Abdalla , C. de Calan and J. A. Mignaco 
for valuable
discussions.
This work was partially supported by CNPq and
FAPESP.

\newpage

\end{document}